\documentclass[12pt,preprint]{aastex}
\shorttitle{The X-ray Jet in the Crab Nebula}
\begin{document}

\title{The X-Ray Jet in the Crab Nebula: Radical Implications for
Pulsar Theory?}

\author{Yury Lyubarsky, David Eichler}

\affil{Department of Physics, Ben-Gurion University, Beer-Sheva
84105, Israel}
 \email{lyub,eichler@bgumail.bgu.ac.il}


\begin{abstract}
The recent Chandra image of the Crab nebula shows a striking,
axisymmetric polar jet.
It is shown that jets are formed in axisymmetric, magnetized pulsar winds and
that the jet luminosity scales relative to the total as
$(\gamma_0\sigma_{eq})^{-4/3}$, where $\sigma_{eq}$ is the ratio of Poynting
flux to particle kinetic energy output at the equator at the base
of the flow and $\gamma_0$ the initial Lorentz factor of the flow. 
The results are applied to the image of
the Crab nebula, and the limit is set for the Crab pulsar of
$\sigma_{eq} \leq 10^2$. It is argued that conventional pulsar
theory needs to be reexamined in light of these limits.
\end{abstract}
\maketitle

\section{Introduction}

Magnetic collimation of astrophysical jets was proposed over 20 years ago 
(Blandford 1976, Lovelace 1976, Benford 1978).
Since then, however, there has been considerable debate as to how
well it works. Heyvaerts and Norman (1989) and Chiueh, Li, and
Begelman (1991) proved that given enough time, all rotating,
axisymmetric magnetized flows are collimated by the toroidal
magnetic field that is inscribed by the rotation. On the other
hand, it was argued (Eichler 1993, Begelman \& Li 1994, Tomimatsu 1994, 
Bogovalov 1997, 2001,
Beskin, Kuznetsova \& Rafikov 1998, Chiueh, Li \& Begelman 1998) that the
collimation is so slow (logarithmic) that in practice it could take
an exponentially large radius to collimate appreciably, so that all
but a small core near the axis would remain uncollimated. Moreover,
if the magnetic field dominates the flow, collimation would come at
the expense of kink instability. The slowness of the collimation
and/or its questionable stability was felt by some
to make it unattractive.

For relativistic outflows with Lorentz factor $\gamma \gg 1$, the
problem is particularly severe. Firstly, the radius
of curvature of the flow lines scales as $1/\gamma^2$, so
that for large enough $\gamma$, the radius of curvature is not only
exponentially large, but exponential large with a huge exponent.
Secondly, even if the flow started at modest $\gamma$, if it were
magnetized, even a small degree of collimation would imply
acceleration (Begelman \& Li 1994) to larger $\gamma$, which places
even more severe limits on collimation.

However, the remarkable image of the Crab Nebula in the X-rays
recently obtained by the CHANDRA Observatory (Weisskopf et al.\ 2000) convincingly shows 
a very well defined polar jet. It appears to be a) very well
collimated but b) only a small part ($\sim 10^{-2} - 10^{-1}$ by
visual inspection) of the total energy, which is mostly expelled in an
apparently equatorial disk. This vindicates both the contentions
that a) magnetic collimation can work and that b) it only works on
the inner core of an ultra-relativistic outflow. Stability worries
may be minimized if most of the jet has been collimated only within
its most recent dynamical timescale.

The question is now whether theory can accurately account for the
proportions of the polar jet relative to the total pulsar wind,
whether the predicted Lorentz factor of the jet is consistent with
observational limits on relativistic beaming, and whether the
implications of the pulsar wind observations are consistent with
the theory of pulsar magnetospheres.

In this paper we consider these questions. We show that the
conventional view of the Crab pulsar would allow too miniscule a
jet to be consistent with the observed one. We then show that
relaxing the usual beliefs about the Crab pulsar and assuming only
that the pair luminosity must be less than the gamma ray luminosity
allows a jet fraction at most $10^{-2}\gamma_0^{-4/3}$, just barely 
consistent with the observations even for $\gamma_0\sim 1$. Here
$\gamma_0$ is the Lorentz factor at the base of the flow.
In the following sections, we show that the jet luminosity scales
relative to the total as $(\gamma_0\sigma_{eq})^{-4/3}$, where 
$\sigma_{eq}$ is the ratio of Poynting flux to particle kinetic energy 
output at
the equator at the base of the flow. We note  in the discussion section
the implications
of this for theories of the Crab pulsar.

\section{Basic equations}
We first review the basic theory of relativistic, magnetized winds.
At large distances (in fact beyond the fast magnetosonic point) one
can consider the flow as a purely poloidal ($v_{\varphi}=0$) flow
in a purely toroidal magnetic field (cf.\ Li 1996). Indeed conservation of the
angular momentum implies that the azimuthal velocity goes to zero
with radius while the contribution of the
poloidal field to the force balance becomes eventually negligible
because the poloidal field falls off faster than the toroidal one.
In ultrarelativistic flows, hoop stress and electrical force nearly
cancel each other so that the poloidal field pressure, while much
less than either term in magnitude, is negligible only far enough
from the light cylinder to be smaller than their difference.
However, this distance still lies well within the termination
shock, and the logarithm of the ratio between the two will be shown
to play a role.
The momentum conservation equation is given by
$$
\rho(\gamma{\bf v}\cdot\nabla)\gamma{\bf v}=\rho_e{\bf E}+{\bf j\times B},
\eqno(1)
$$
where $\rho$ is the proper mass density, $\bf v$ the plasma velocity,
$\bf E$ and $\bf B$ are
the electric and magnetic field strengths, correspondingly and
$\rho_e$ and $\bf j$ are the charge and current densities,
correspondingly. Under the above assumptions one can write
$$
{\bf v}=v{\bf l};\qquad {\bf B}=B{\bf e}_{\varphi};\qquad {\bf
E=-v\times B}=-vB{\bf t};
\eqno(2)
$$
where ${\bf l}$ is the
longitudinal unit vector along the magnetic surface,
$e_{\varphi}$ is the azimuthal unit vector and
${\bf t=l\times e}_{\varphi}$ is the unit vector in the transverse
direction. Ee take the speed of light to be
unity. With the Maxwell equations, we can present the current
and charge densities in the form
$$
4\pi{\bf j}=\nabla\times {\bf
B}=\frac1r\nabla(rB){\bf\times e}_{\varphi};
\eqno(3)
$$$$
4\pi\rho_e=\nabla\cdot{\bf [B\times v]}=-\frac vr{\bf
t\cdot\nabla}(rB)- B{\bf t\cdot\nabla}v-vB({\bf
e}_{\varphi}{\bf\cdot[\nabla\times l]}),$$ where $r$ is the
cylindrical radius.

The cross-field equation may be obtained by taking the dot product
of Eq.(1) with $\bf t$ and using the above expressions for $\rho_e$
and $j$. After simple algebra, we obtain
$$ 
\rho(\gamma^2-1){\bf t\cdot(l\cdot\nabla )l}=\frac{1}{4\pi}
\left\{ v^2B^2({\bf e}_{\varphi}{\bf\cdot[\nabla\times l]})-
\frac{B}{r\gamma}{\bf t\cdot\nabla} \frac{rB}{\gamma}\right\}, 
\eqno(4) 
$$ 
where $\gamma=(1-v^2)^{-1/2}$ is the flow Lorentz factor. Taking into account
the definition of the curvature radius of the flux line 
$$
\frac{1}{R_c}\equiv {\bf t\cdot(l\cdot\nabla )l}=-{\bf
t\cdot[l\times[\nabla\times l]]}= -{\bf
e}_{\varphi}{\bf\cdot[\nabla\times l]}, 
$$ 
one can write finally 
$$
\frac{\gamma^2-1}{R_c}\left(1+\frac{1}{\sigma}\right)=-{\bf
t\cdot\nabla}(\ln\frac{rB}{\gamma}), 
\eqno(5) 
$$ 
where the ratio of the Poynting flux to the kinetic energy flux is 
$$
\sigma=\frac{EB}{4\pi\rho v\gamma^2}=\frac{B^2}{4\pi\rho\gamma^2}.
\eqno(6) $$
One can see from Eq.(5) that in an ultrarelativistic wind, $\gamma\gg
1$, the flow lines are nearly straight and collimation is possible, if
possible at all, only near the axis (Tomimatsu 1994,
Chiueh et al. 1998, Bogovalov 2001).

The Lorentz factor of the flow should be determined from the energy
conservation along the flux tube. This equation may be obtained by
taking the dot product of the momentum equation (1) with the
longitudinal vector $\bf l$. However one can instead simply write
down the energy flux
per unit poloidal flux 
$$
\frac{\rho\gamma^2v+\frac{EB}{4\pi}}{B_p}=\it const. 
$$ 
Taking into
account conservation of the mass flow within the field line tube,
we define a quantity $\eta(\Psi)$ that is conserved along each
surface $\Psi$ by 
$$ \rho v\gamma\equiv\eta(\Psi)B_p. 
\eqno(7) 
$$
Making use of this equation and the relation 
$$ E=r\Omega B_p,
\eqno(8) 
$$ 
one can write the energy equation as 
$$ 
\gamma
+\frac{r^2\Omega^2 B_p}{4\pi\eta (\Psi)v}\equiv\mu(\Psi). \eqno(9)
$$ Here the poloidal flux function is defined from the poloidal
field strength by $$ {\bf B}_p=\frac{1}{r}\, \nabla\Psi\times{\bf
e}_{\varphi}. $$  
It follows
from Eq.(9) that plasma accelerates if
$r\vert\nabla\Psi\vert$ decreases (Begelman \& Li 1994). In both
radial flows and cylindrical ones the Lorentz factor remains
constant. (Note that in the case of interest one need not introduce
the poloidal field at all but instead define $\Psi$ as a stream
function (cf.\ Contopoulos 1995). We retained the standard
definitions to provide a simple comparison with the previous works.)

The asymptotic transfield and energy equations were found by expanding the
full set of MHD 
equations in $1/r$ by 
Begelman \& Li (1994), Tomimatsu (1994), Bogovalov (1997, 2001), Chiueh et al
 (1998), Bogovalov \& Tsinganos (1999). One can show that their equations are 
equivalent to Eqs.(5, 9). Examining
the neglected terms one can see that these equations are valid at $r\gg\gamma R_L$
where $R_L=c/\Omega$ is the light cylinder radius and $\Omega$ the angular velocity of 
the pulsar. This condition arises because the neglected poloidal field stress 
$(\propto B_p^2\propto (r/R_L)^4$) should be compared in the relativistic case not 
with the hoop stress 
($\propto B^2_{\varphi}\propto (r/R_L)^2$) but with $B^2_{\varphi}-E^2=B^2_{\varphi}/\gamma^2$. 
This condition is equivalent to the condition $R\gg R_f$, where $R_f$ is the 
radius of the fast magnetosonic point (Beskin et al. 1998).

\section{The flow near the axis}
Close enough to the axis, a jet may be formed if flow lines turn
through even a small angle $\Delta\theta\sim\theta$. Let us
consider the flow near the axis. To specify the conserving fluxes
one should know the flow structure in the near zone.
Note that at distances exceeding $R_L$ the flow fills all the space
and plasma density and other parameters vary at the angular scale
$\theta\sim 1$. One can naturally assume that close enough to the
axis, $\theta\ll 1$, both plasma density and the initial Lorentz factor are
close to the values at the axis. Therefore we can take as a good approximation 
$\eta(\Psi)=\it const$ and $\gamma_0=\it const$. The flow lines of the
pulsar wind are practically straight inside the fast magnetosonic point
(but outside the light cylinder) and the flow is nearly quasiradial
in this region. An example of such a flow is provided by the well-known
split monopole solution (Michel 1973)
 $$
\Psi=\Psi_{eq}(1-\cos\theta)\approx\frac12\Psi_{eq}\theta^2,\qquad
B_{\varphi}=\frac{\Psi_{eq}\Omega\sin\theta}{cR},\qquad {\bf
E}=\frac{\Omega}c\nabla\Psi. 
\eqno(10) 
$$
Here $R$ and $\theta$ are
spherical radius and polar angle, correspondingly, and $\Psi_{eq}$
is the total (equatorial) poloidal flux. This solution is valid within
the fast magnetosonic point. Taking into account that
$\Psi\propto\theta^2$ at $\theta\ll 1$ in any quasiradial flow, one
can describe any such a flow near the axis by an equivalent split monopole
solution. Note that even in flows that are eventually collimated (e.g., Eichler 1993)
the outflows well within the collimation scale reduce to (10), which
represents the precollimated state. 

The energy integral $\mu$ should in principle be determined by the
conditions at the fast magnetosonic point. Careful considerations show
that deviations from quasiradial propagation remain small at this point
in high $\sigma$ flows (Begelman \& Li 1994, Tomimatsu 1994, Beskin et al
1998). Though these deviations are crusial for determining
the position of this point, one can safely neglect them
 in the angular distribution of the parameters. Substituting equation (10)
 into equation (9), one gets
$$
\mu=\gamma_0+\frac{\Omega^2\Psi_{eq}\sin^2\theta}{4\pi\eta v_0}.
$$
Expressing back $\theta$ via $\Psi$, we find the approximate energy
integral as
$$ 
\mu (\Psi)=\gamma_0 \left\{1+
\sigma_{eq}\left(2\Psi/\Psi_{eq}-(\Psi/\Psi_{eq})^2\right)\right\}\approx
\gamma_0\left(1+2\sigma_{eq}\Psi/\Psi_{eq}\right),
\eqno(11) 
$$ 
where $\sigma_{eq}$ is the ratio of the Poynting to the matter energy
flux at the equator. Because we are only interested in the flow
near the axis, the "equatorial" values should be considered not as
true values at the equator of the flow but as values at the equator
of the split monopole flow that coincides close to the axis with
the flow of interest.

From here on we will use the small angle approximation
$\psi\equiv\Psi /\Psi_{eq}\ll 1$, $r\ll z$ etc. We will consider
the flow as ultrarelativistic, $\gamma\gg 1$, because a Poynting
dominated flow beyond the fast magnetosonic point is necessarily
ultrarelativistic. Note that Poynting flux decreases toward the
axis, the initial ratio of the Poynting flux to the kinetic energy
flux being $$ \sigma_0=2\sigma_{eq}\psi. \eqno(12) $$ Thus, close
enough to the axis, $\psi<(2\sigma_{eq})^{-1}$, the flow is
matter dominated. This part of the flow is collimated already at
the distance $R\sim R_f$ (Bogovalov 2001). We are interested in
collimation of a wider flux tube where the flow is Poynting
dominated. 

Note that decreasing of the Poynting flux toward the axis may be understood
from a rather simple consideration that demonstrates that this feature is
quite general and not a specific feature of Michel's solution. It follows
immediately from equation (8) that the electric field decreases toward the 
axis. The toroidal magnetic field $B_{\varphi}\propto r^{-1}\int jrdr$ also
decreases unless the current density $j$ is singular at the axis. Therefore
the Poynting flux $EB_{\varphi}/4\pi$ decreases toward the axis. Our
expression (11) for $\mu(\psi)$ may be considered simply as expansion
of general energy flux in small $\psi$.

The shape of the flux surfaces is conveniently described
in cylindrical coordinates by the function $r(\psi,z)$ instead of
$\psi(r,z)$. Then, e.g., $$ B_p=\frac1r\vert\nabla\psi\vert\approx
\frac1r\frac{\partial\psi}{\partial r}= \left(r\frac{\partial
r}{\partial\psi}\right)^{-1}. $$
In the small angle approximation the curvature radius may be presented as
$$ R_c=\left(\frac{\partial^2r}{\partial z^2}\right)^{-1}. $$ It
follows immediately from Eqs (6, 8) that $\sigma\propto rB/\gamma$
when $\eta(\psi)=\it const$. Transforming derivatives to
independent variables $r$ and $\psi$,
one can write Eq.(5) as
$$ \gamma^2\frac{\partial
r}{\partial\psi}\frac{\partial^2r}{\partial z^2}=
-\frac{\partial\ln(1+\sigma)}{\partial\psi}. \eqno(13) $$ The
energy equation (9) is now written, with the aid of Eq.(11), as 
$$
\gamma+\frac{\sigma_{eq}\gamma_0}v r\vert\nabla\psi\vert=
\gamma_0 (1+2\sigma_{eq}\psi).
$$
Taking into account that $\gamma^3\gg\gamma_0\sigma_0$ beyond the fast 
magnetosonic point (see, e.g., Beskin et al.\ 1998), one can substitute
$v=1$ into the denominator of the second term in the left-hand side of this
equation and write finally
$$
\gamma= \gamma_0\left[1+\sigma_{eq}\left(2\psi-
r\left(\frac{\partial r}{\partial\psi}\right)^{-1}\right)\right].
\eqno(14) 
$$ The system of equations is completed by the expression
for $\sigma$ (see Eqs.(6, 7, 12)): 
$$
\sigma=\frac{\Omega^2r^2B_p}{4\pi\eta\gamma}=
\sigma_{eq}r\vert\nabla\psi\vert\frac{\gamma_0}{\gamma}=
\sigma_{eq}r\frac{\gamma_0}{\gamma}\left(\frac{\partial
r}{\partial\psi}\right)^{-1}. 
\eqno(15) 
$$

\section{Collimation of the flow}
Bogovalov (2001) very recently showed numerically that in the case of 
$\gamma_0^2\le\sigma_{eq}$
the flow within the tube $\psi<(2\sigma_{eq})^{-1}$ is
collimated into a cylindrical jet at the distance of the order of the radius of the 
fast magnetosonic point. One can see that within this flux tube
$\sigma_0\le 1$ and the fraction of the energy carried by the jet is only 
$\sim\sigma_{eq}^{-2}$.
This is insufficient to explain the observed jet in the Crab nebula.
Here we consider whether a larger fraction of the poloidal flux (and, 
correspondingly,
of the total energy) may collimate at larger radii.

The initial radial flow (10) may be presented near the axis as
$r=\sqrt{2\psi}z$. Let us look for small deviations from this
radial flow, $$ r=\sqrt{2\psi}z(1-y);\qquad y(\psi,z)\ll 1. $$ The
approximate collimation condition will be defined as follows. We
shall solve the equations retaining only terms of the lowest order
in $y$ and if we find $y\to 1$ for
some poloidal field line, we conclude that this line collimates.

\subsection{The case $\psi<(2\sigma_{eq})^{-1}$}
It follows from Eq.(11) that under the condition
$\psi<(2\sigma_{eq})^{-1}$
the flow is matter dominated.
One can approximate this case by assuming $\sigma\ll 1$, which is
formally correct for $\psi\ll (2\sigma_{eq})^{-1}$. In this case
the flow evidently does not accelerate, $\gamma\approx\gamma_0$,
and $\sigma$ varies insignificantly, $\sigma\approx\sigma_0$, as
long as deviation of the flow from the purely radial one is small,
$y\ll 1$. Linearizing Eq.(13), one gets with the aid of Eq.(12) 
$$
z\gamma_0^2\frac{\partial^2 (zy)}{\partial z^2}=2\sigma_{eq}. 
$$ The
solution to this equation is 
$$
y=\frac{2\sigma_{eq}}{\gamma_0^2}\ln\frac{z}{R_f}. 
$$ 
The initial
point was chosen to be $R_f$ because the equations are valid only
at $z\gg R_f$ and there is no collimation at $z\ll R_f$ (Begelman \& Li 1994). 
One can see that if 
$$
2\sigma_{eq}<\frac{\gamma_0^2}{\ln\frac{R}{R_f}}, 
\eqno(16) 
$$
where $R$ is the radius of the termination shock, the flow is not
collimated at all. If the reverse is true, i.e. if 
$$
2\sigma_{eq}>\frac{\gamma_0^2}{\ln\frac{R}{R_f}}, 
\eqno(17) 
$$ 
the flow collimates at $\psi<(2\sigma_{eq})^{-1}$. At
$2\sigma_{eq}\ge\gamma_0^2$ the flow tube $\psi<(2\sigma_{eq})^{-1}$
collimates already at $z\sim R_f$, which agrees with the results by
Bogovalov (2001). At the condition (17) more flow than this may
collimate but as will be seen below, such additional collimation
invariably introduces acceleration, thus invalidating the
approximation of the constant $\gamma$.

\subsection{The case $\psi>(2\sigma_{eq})^{-1}$, no acceleration}
In this region we can take $\sigma\gg 1$. Let us first assume that
the flow is not accelerated significantly and show that this
implies no further collimation. Substituting $\gamma=\gamma_0$ and
$\sigma=\sigma_0$ into Eq.(13) and linearizing it, one gets $$
z\psi\frac{\partial^2(zy)}{\partial z^2}=\frac{1}{\gamma^2_0}.
\eqno(18) $$ The solution to this equation is $$
y=\frac{\ln\frac{z}{R_f}}{\gamma_0^2\psi}. \eqno(19) $$

Because in the region of interest $\sigma_0\gg 1$ even a small
deviation from the radial flow may result in significant growth of
$\gamma$. The linearized energy equation (14) reads 
$$ \gamma=
\gamma_0\left(1-4\sigma_{eq}\psi^2\frac{\partial
y}{\partial\psi}\right). 
\eqno(20) 
$$ 
Our approximation $\gamma\approx\gamma_0$ 
is valid as long as the second term in the
brackets remains small as compared with unity. Substituting the
solution (19), one can see that our assumption of constant $\gamma$
is already violated at $R\sim R_f$ if $4\sigma_{eq}\ge\gamma_0^2$;
this case will be considered in the next subsection. By Eq.(20) the
solution (19) is valid at $$
4\frac{\sigma_{eq}}{\gamma_0^2}\ln\frac{z}{R_f}<1. \eqno(21) $$ If
it is valid until the termination shock, $y$ remains less than
unity for any $\psi>(2\sigma_{eq})^{-1}$, so we come back to
condition (16). If the reverse is true, the acceleration becomes
significant at 
$$
z_0=R_f\exp\left(\frac{\gamma_0^2}{4\sigma_{eq}}\right)<R.
\eqno(22) 
$$ At this point $y\ll 1$. However, beyond this point
collimation may ensue. So let us consider collimation in the
accelerating flow.

\subsection{The case $\psi>(2\sigma_{eq})^{-1}$, accelerating flow}
Now one can neglect the first term in Eq.(20) and, substituting the
 Lorentz factor so obtained into Eqs.(13, 15) and neglecting the
higher order terms in $y$, one obtains
 $$
-16\gamma_0^2\sigma_{eq}^2\psi^5z\left(\frac{\partial
y}{\partial\psi}\right)^3 \frac{\partial^2 (zy)}{\partial z^2}=
\frac{\partial y}{\partial\psi}+\psi\frac{\partial^2
y}{\partial\psi^2}. 
\eqno(23) 
$$ 
The solution to this equation with
logarithmic accuracy (i.e.\ neglecting terms of $[\ln
(R/R_f)]^{-1}$) is $$ y=\frac{\left((3/2)\ln\frac
z{R_0}\right)^{1/3}}{2\psi(\gamma_0\sigma_{eq})^{2/3}}. \eqno(24)
$$ Here $R_0=R_f$ if $4\sigma_{eq}\ge\gamma_0^2$ and $R_0=z_0$ if
$\left(\ln\frac{R}{R_f}\right)^{-1}<4\sigma_{eq}/\gamma_0^2<1$. The
fraction of the collimated flux may be estimated from the condition
$y\sim 1$ as 
$$
\psi_c=\frac{\left((3/2)\ln\frac{R}{R_0}\right)^{1/3}}{2(\gamma_0\sigma_{eq})^
{2/3}}.
\eqno(25) $$

\subsection{Energy considerations}
Substitution of Eq.(24) into Eq.(20) yields 
$$
\gamma=2\left(\frac32\gamma_0\sigma_{eq}\ln\frac{z}{R_0}\right)^
{1/3},
\eqno(26) 
$$ 
which is consistent with the results by Tomimatsu
(1994) and Beskin et al. (1998). In principle, the flow may
accelerate up to the Lorentz factor $\gamma_{max}(\psi)
=\gamma_0\sigma_0=2\gamma_0\sigma_{eq}\psi$. At the collimation
point defined from Eq.(24) as a point where $y\sim 1$, one obtains
$\gamma\sim\gamma_{max}(\psi)$. So when a flux line turns
through an angle $\Delta\theta\sim\theta$, a significant fraction
of the Poynting flux is transferred to the plasma and there should
be $\sigma\sim 1$ in the collimated part of the flow. The
collimation (and acceleration) process has enough time to be nearly completed
only within the small fraction (25) of the poloidal flux.
Outside this flux tube the flow slowly accelerates according to
Eq.(26). However, when the flow enters the termination shock, most
of the energy still remains in the form of Poynting flux.

Because Poynting flux decreases toward the axis of the flow, the
fraction of the energy carried by the jet is small. Integrating the
energy flux (11), one gets $\dot E_{jet}/\dot E_{tot}=1.5\psi_c^2$.
So the jet carries the fraction of the total energy $$ \frac{\dot
E_{jet}}{\dot E_{tot}}=
\frac{3\left((3/2)\ln\frac{R}{R_0}\right)^{2/3}}{8(\gamma_0\sigma_{eq})^{4/3}}.
\eqno(27) $$

\section{Discussion}
The fraction of the collimated flux is determined by the parameter
$\gamma_{max}=\gamma_0\sigma_{eq}$, which is Michel's (1969) magnetization
parameter.
This parameter is determined by the amount of plasma emitted by the
pulsar. The density of
the plasma generated within
the pulsar magnetosphere is conveniently normalized by the Goldreich-Julian
charge density:
$$
n=\frac{\kappa\rho_{GJ}}e=\frac{\kappa B\Omega}{2\pi ec},
$$
where $B$ is the magnetic field in the magnetosphere, $\kappa$ the multiplicity
factor. Within the light cylinder the field
is nearly dipolar, $B=\mu/R^3$, where $\mu$ is the magnetic moment of the
star, and
conservation of the mass flux implies $n\propto B$. Beyond the light
cylinder $n\propto 1/R^2$.
The Poynting flux in the pulsar wind may be expressed via the magnetic
field at the
light cylinder, $B_L=\mu\Omega^3/c^3$, as
$$
W=\frac{cB_L^2}{4\pi}\left(\frac{R_L}R\right)^2.
$$
Now one can estimate the magnetization parameter as
$$
\gamma_{max}=\frac W{mnc^3}=\frac{eB_L}{2mc\kappa\Omega}=
1.3\times 10^3\frac{\mu_{30}}{\kappa_4P^2},
\eqno(28)
$$
where $\mu =10^{30}\mu_{30}$ G$\cdot$cm$^3$, $\kappa=10^4\kappa_4$ and $P$
is in seconds.
For the Crab pulsar, where $P=0.033$ s and $\mu_{30}=5$, we have
$$\gamma_{max}=\frac{6\cdot 10^6}{\kappa_4}.\eqno(29)$$
From the observed amount of radio emitting electrons (those with the energy $>100$MeV)
one can get an upper limit for the injection rate of these electrons about
$\sim 10^{41}$ s$^{-1}$, which corresponds to $\kappa\sim 3\cdot 10^6$.
If the pulsar ejects a larger amount of pairs, their energy in the nebula
should be less than $100$ MeV.

The standard models of plasma production in pulsars (see, e.g., the
review by Arons 1983) typically predict $\kappa\sim 10^3-10^4$ and
$\gamma_0\sim 100$. Different modifications
were proposed to explain the observed powerful gamma radiation
(Cheng, Ho \& Ruderman 1986, Usov \& Melrose 1996, Lyubarskii
1996); these models predict significantly larger plasma production.
However, we are not aware of models predicting $\kappa>{\rm
few}\times 10^6$. If protons are also emitted by the pulsar, the
corresponding magnetization parameter is reduced by a factor
$m/m_p$. However theoretical arguments (Arons 1983) favor
$\kappa_{p}\sim 1$.

It is very difficult to place limits on the wind parameters directly from
observations.
Limitations found by Wilson \& Rees (1978) who studied induced scattering
of the 
radio emission in the pulsar wind were obtained on the assumption that the
cyclotron
frequency near the light cylinder is less than the radiation frequency.
However the reverse
is true in the case of the Crab pulsar (Barnard 1986, Lyubarskii \& Petrova
1996). The
scattering rate in the magnetized plasma is small if the pulsar radiates
only waves polarized
perpendicularly to the local magnetic field (Blandford \& Scharlemann 1976,
Lyubarskii \&
Petrova 1996). In this case the scattering begins only beyond the
polarization limiting radius,
which lies, in the Crab case, well outside the light cylinder.

Because pairs are produced by gamma radiation, one can reasonably conclude
that the pair
luminosity does not exceed the gamma luminosity. This places the
limit $\sigma_{eq}>100$. One can see from Eq.(27) that $\dot
E_{jet}/\dot E_{tot}\approx 10^{-2}$ even at $\sigma_{eq}=100$ and
$\gamma_0=1$.

Although the fraction of total luminosity that
emanates from the jet is not yet published, and in any case is not
quite the same as the fraction of mechanical output that emerges in
the jet, visual inspection seems to imply that $10^{-2}$ is a
conservative guess. The implication that $\sigma_{eq}$ is {\it at
most} $10^2$ and $\gamma_0\sim 1$ runs counter to conventional 
pulsar wisdom. Probably observed constraints on relativistic beaming
will provide an even sharper constraint on $\sigma_{eq}$ and $\gamma_0$,
making the observed jet even more confusing for conventional pulsar theory.
However, since existing theory for pulsar
magnetospheres is not rigorous, perhaps it needs to be reexamined. The jet may be formed 
via hoop stress collimation of the the pulsar wind only if the wind is much more mass-loaded
 than previously believed.
The alternative possibility, that the jet is formed already near or within the pulsar magnetosphere,
$R\sim R_L$, also requires serious modifications to the conventional picture of the
pulsar magnetosphere.

We thank the Israeli Science Foundation and the Arnow Chair of Physics
for their generous support.

\end{document}